# Detailed-balance efficiency limits of two-terminal perovskite/silicon tandem solar cells with planar and Lambertian spectral splitters

Verena Neder[1,2], Stefan W. Tabernig[2], Albert Polman[2]

[1]Institute of Physics, University of Amsterdam
Science Park 904, 1098 XH Amsterdam, the Netherlands

[2]Center for Nanophotonics, NWO-Institute AMOLF
Science Park 104, 1098 XG, Amsterdam, the Netherlands

**Abstract**

We derive the photovoltaic conversion efficiency limit for two-terminal tandem solar cells with a perovskite top cell and silicon bottom cell with an embedded spectrum splitter. For large-bandgap top-cells a spectrum splitter strongly enhances the efficiency because of enhanced light absorption and trapping. A Lambertian spectral splitter shows a significantly improved effect compared to a planar splitter: we find an ideal efficiency enhancement in the thermodynamic limit for a 500 nm thick top cell of 6% absolute for bandgaps above 1.75 eV. Vice versa, the use of a spectral splitter geometry enables the use of a thinner top cell. Using experimental parameters for perovskite cells we show that for a top-cell bandgap of 1.77 eV a 2.8% absolute efficiency can be gained. The calculations in this work show that integration of a spectral splitter into perovskite/silicon tandem cells with a top bandgap above 1.7 eV can lead to a large increase in efficiency, even with realistic experimental losses and non-unity reflection of the spectral splitter.

**Introduction**

At present, solar photovoltaics (PV) has an installed capacity of around 600 GW$_p$ worldwide [1]. Following the IRENA roadmap, by 2050 PV should reach an installed capacity of 8.5 TW$_p$ to account for 43% of the total installed power capacity for electricity generation [2]. With the costs of the cells determining only a small part of the costs of a PV system, raising their efficiency is a key method to reduce the cost of PV per kW$_p$. Also, higher-efficiency panels take up less space, which is essential as PV is applied at very large scale. Therefore, in PV research it is crucial to fight for every digit that can be gained in cell efficiency. With 95% of the total production in 2019, the market is strongly dominated by single-junction Si-wafer based PV technology [3]. Silicon-based tandem solar cells have the potential to raise the efficiency beyond the theoretical limit of 29.4% for Si-only cells [4]. The combination of perovskite and Si in a 2-terminal (2T) or 4-terminal (4T) tandem configuration is one of the most growing and promising concepts. Recently, a Si-perovskite 2T tandem solar cell was presented with an efficiency of 29.15% [5] well above the record for a single-junction Si cell of 26.7% [6].

One aspect that is of high importance to improve the performance of tandem cells is light management to optimize the coupling and distribution of sunlight in the tandem subcells [7]–[10]. Recent work has focused on minimizing reflection from the top or interlayers [11]–[14] and reducing parasitic absorption in the inactive layers such as transparent conductive layers [9], [15], [16], and to optimize light trapping in the top and bottom cells [17], [18]. A concept that has not been studied in much detail is to control the spectral splitting of light directed into the two subcells [19]. Spectral splitting can be achieved by an additional interlayer between the top- and bottom cells that effectively reflects the part of the spectrum with energy above the bandgap of the top cell, and that transmits the remainder to the bottom cell. In an ideal case, the low- and high-energy spectral bands are fully split between the cells, so that maximum current and voltage can be harvested. However, in practical geometries a low-energy tail close to the top-cell bandgap is always transmitted due to incomplete light absorption in the top cell, and is then absorbed in the underlying cell (Figure 1a). This transmitted tail creates higher thermalization losses in the bottom cell and should thus be avoided.

Earlier, spectral splitting in 4T tandem cells has been modelled [19] and it was predicted that, by integrating a spectral splitting light trapping layer, an efficiency gain between 0.5% and 3% (absolute) and a two- to threefold reduction in thickness of the top cell can be reached, depending on the diffusion length of the absorber material. The benefits of spectral splitting in 2T tandem concepts have been studied previously to achieve current matching between top and bottom cell, using Bragg reflectors [20]–[23] or three-dimensional photonic crystals [24]–[26] as intermediate layers. Current matching is a limiting factor in 2T tandems, in particular for top cell materials with a bandgap below $E_{BG}$=1.73 eV. In that case the current generated in the perovskite can exceed that in the Si, depending on the absorption in the top cell, in which case no spectral splitter is needed. However, a spectral splitter enables the use of a smaller perovskite cell thickness to obtain current matching with low-gap perovskites. In addition, for tandems with higher perovskite bandgaps, which have the largest potential tandem efficiency, and which have applications in photoelectrochemical splitting of water because of their high voltage [27], [28], for example, spectrum splitting has significant potential to enhance the efficiency. In the following, we show for which conditions a spectral splitter in a 2T perovskite/Si tandem cell is beneficial and what efficiency gains can be achieved. We make a distinction between planar spectral splitters with specular reflectivity, and spectral splitters with a Lambertian scattering distribution that enhances light trapping in the perovskite top cell.

**Methods**

Throughout the article, we use detailed-balance calculations [29] to determine the theoretical maximum efficiencies of the different 2T tandem configurations. Auger recombination in Si and other non-radiative processes are not taken into account and we assume full current collection from light that is absorbed in the top- and bottom cells. While such theoretical maximum efficiencies cannot be reached in reality, they allow a reliable comparison between different configurations. Similar trends as we find here will also apply for tandem designs in realistic experimental geometries. We differentiate between three cases of light absorption in the perovskite top cell:

1. Full absorption up to the bandgap as is typically done in detailed-balance limit calculations;

2. Ideal current splitting that assumes that the top cell absorbs exactly the amount of light such that half of the maximum possible current of the bottom cell (as a single-junction cell) is generated in the top cell;

3. Single pass absorption of a semi-transparent top cell with finite thickness *d*:

$$A_{TTC} = (1 - \exp(-\alpha \cdot d)) \qquad (1)$$

with α the absorption coefficient for the transmitting top cell (TTC). To model the bandgap-dependent absorption in the perovskite top cell we use the wavelength-dependent absorption coefficient of

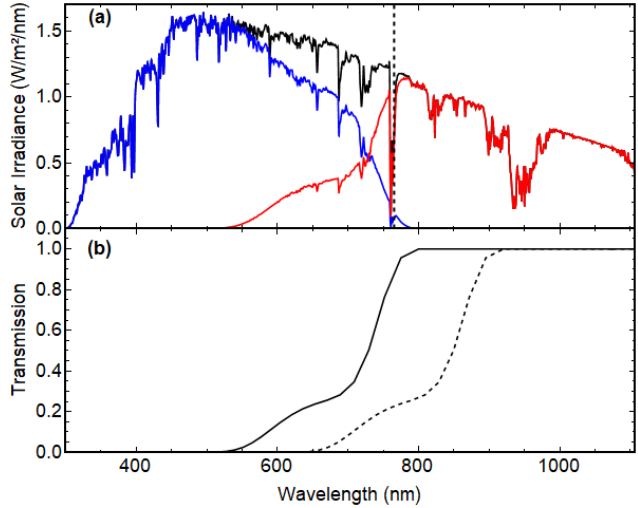

**Figure 1** (a) AM 1.5G solar spectrum (black line) with theoretical absorption in 500 nm thick perovskite top cell (blue line) and spectrum directed into Si bottom cell (red line). The vertical dashed line marks the position of the bandgap $E_{BG}$ = 1.62 eV. (b) Transmission spectrum of 500 nm thick perovskite with $E_{BG}$ = 1.62 eV (solid black line) from [30] and modelled transmission for 500 nm thick perovskite with $E_{BG}$ = 1.4 eV (dashed black line).

CsFAPbIBr with a bandgap of $E_{BG}$ = 1.62 eV [30] which is in the bandgap range of the most frequently used perovskite top cells. Figure 1b shows the transmission of a 500 nm thick layer of this material (black solid line). The transmission for perovskites with the same thickness but other bandgap was then modelled by shifting the graph by the bandgap shift (Fig. 1b). Our simplified model of light absorption might not fully reflect transmission for specific perovskites with different material compositions and bandgaps, however, it allows us to systematically compare the effect of transparency of the top cell for different bandgaps.

**2T tandem efficiencies with semitransparent top cells**

The thermodynamic (TD) limiting efficiency for 2T tandems, assuming perfect absorption up to the bandgap in the top cell of the perovskite/Si tandem, is maximal for $E_{BG}$=1.73 eV with an efficiency of 45.1%. [31] (Fig. 2). The limiting efficiency gradually decreases as the top-cell bandgap approaches the Si bandgap (1.12 eV), where all light up to that energy is absorbed in the top cell with no current left for the bottom cell. We then define the current splitting (CS) limit as a condition, for a given perovskite bandgap, where the top cell is not fully absorbing up to its bandgap, but rather absorbs an optimized smaller fraction to obtain current matching (Fig. 2). In this ideal case the top cell absorbs light such that exactly half of the maximum possible current of the bottom cell (as a single-junction cell) is generated in the top cell. This requires a top cell with a bandgap below 1.73 eV. For bandgaps below 1.73 eV the CS limit is well above that for the thermodynamic limit of fully absorbing top cells. For higher perovskite bandgaps full absorption in the top cell is always optimal and equal to the CS limit.

As described above, for any perovskite bandgap the optimized absorption, or equivalently transmission, of the top cells can be realized by selecting the proper top cell thickness (see method section). We calculate the detailed-balance limiting efficiency using the modeled transmission (Eqn. 1) for thicknesses in the range 250-1000 nm as a function of bandgap of the top cell (Fig. 2). Several trends can be observed in this figure. First, the shape of every individual TTC limit is similar to the shape of the thermodynamic limit, with a pronounced maximum efficiency that is quickly decreasing for lower or higher bandgaps. This is because of the large sensitivity of 2T tandem cells to unbalanced currents in the top and bottom cells. For bandgaps below 1.73 eV the maximum efficiency for every thickness is much higher than the one derived from the thermodynamic limit assuming full absorption. The maximum of all four graphs for different thicknesses match the CS curve at the bandgaps where the currents are matched.

Figure 2 shows there is a wide range of perovskite bandgaps (1.5-1.7 eV) for which very high 2T tandem efficiencies can be reached by optimizing the perovskite thickness in the practical range of 250-1000 nm.

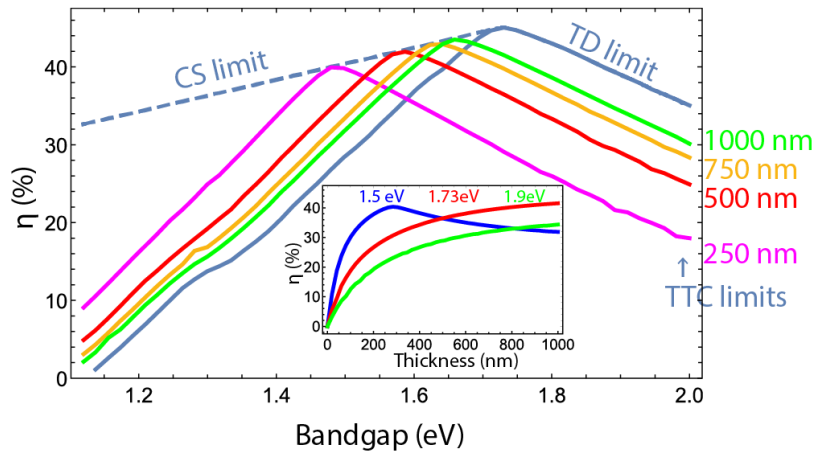

**Figure 2** Detailed-balance efficiency limits for 2T tandem solar cells with Si bottom cell: Thermodynamic limit (TD) assuming perfect absorption in the top cell (solid blue line); Current splitting (CS) limit assuming ideal splitting of the spectrum for perfect current matching in top and bottom cell (dashed blue line); Transmitting top cell (TTC) limit assuming realistic absorption/transmission in a perovskite top cell for top cell thickness in the range 250-1000 nm. Inset: Comparison of dependence of efficiency on top cell thickness for three different perovskite top cell materials with $E_{BG}$ = 1.5 eV, $E_{BG}$ = 1.73 eV and $E_{BG}$ = 1.9 eV.

For top cells with bandgaps higher than $E_{BG}$=1.73 eV, the TTC limit is much lower than the thermodynamic limit, because incomplete absorption in the top cell due to the limited thickness limits the overall current. This is where a spectral splitting light trapping layer could be of special interest. Transmitted light above the bandgap is then reflected and trapped in the top cell, such that the thermodynamic limit could be approached for those bandgaps even with finite top cell thicknesses. This is discussed in detail in the next section.

The described trends are illustrated by the inset in Figure 2 where the TTC limit versus thickness of the top cell is plotted for three different bandgaps. For a bandgap below $E_{BG}$ =1.73 eV an ideal thickness can be chosen to reach the maximum TTC efficiency for that bandgap (e.g. 280 nm at 1.5 eV, see the inset). For bandgaps of $E_{BG}$ =1.73 eV and higher, the cell ideally is 'infinitely' thick to absorb all light above the bandgap energy.

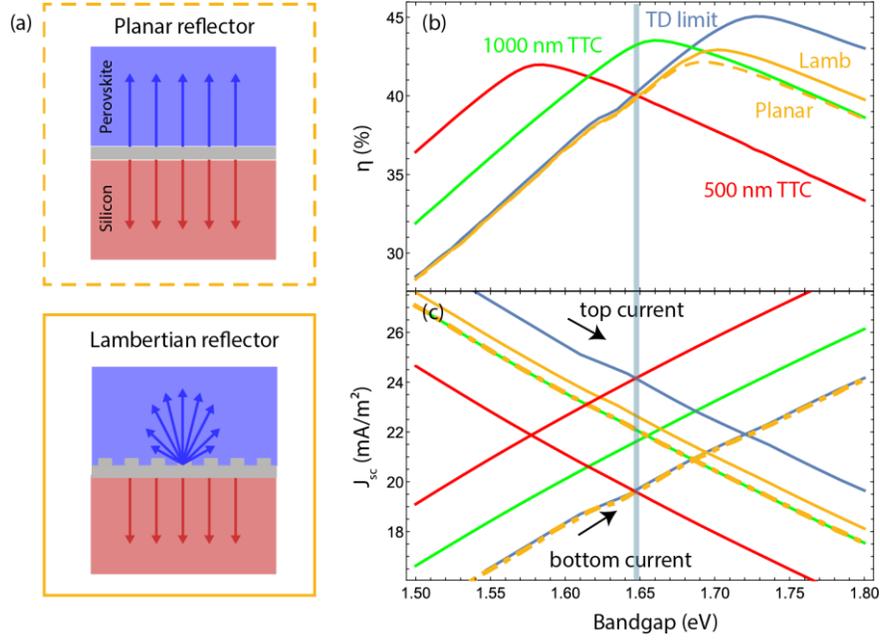

**Figure 3** Spectral splitter in 2T tandem cell. (a) Schematics of planar and Lambertian spectral splitter (in gray). Light is selectively scattered back to the perovskite top cell (blue). The remainder of the spectrum is transmitted to the underlying Si cell (red). (b) Thermodynamic limiting efficiency for full absorption in the top cell (blue line), TTC limit with 500 nm (red) and 1000 nm (green) thick perovskite top cell. Limiting efficiencies of 2T tandems with 500 nm thick perovskite top cell and planar (dashed yellow line) and Lambertian spectral splitter (solid yellow line). (c) J$_{sc}$ of top and bottom cell of tandem cells for all limits from (b). The graphs decreasing and increasing with bandgap represent the current of the top and bottom cells, respectively.

**Planar and Lambertian spectral splitter**

Next we discuss the effect of a spectral splitter in between the top and bottom cell of a 2T tandem cell. We distinguish between: (1) a planar spectral splitter that reflects the light specularly back to the top cell and creates one extra path for absorption in the top cell; (2) a Lambertian spectral splitter that reflects the light back in a cosine angular fashion such that enhanced light trapping can be achieved in the top cell. Figure 3a shows a schematic of both spectral splitter geometries. The reflectivity R(λ) of the spectral splitter is defined as a step function with tunable step wavelength $\lambda_{step}$ and reflectance $r$:

$$R(\lambda) = \begin{cases} 0, \lambda > \lambda_{step} \\ r, \lambda \leq \lambda_{step} \end{cases} \quad (2)$$

In both cases, we assume that light that is not reflected is transmitted losslessly to the underlying Si substrate. We add one additional absorption path in the top perovskite cell with length $d$ for the planar splitter and length $d_{eff} = d \cdot \frac{2+x}{1-x}$ with $x = a(\alpha d)^b$, $a = 0.935$ and $b = 0.67$ to represent the angular distribution scattered by the Lambertian splitter [19], [32]. The absorption in the bottom cell is modified to account for the reflectivity $R$.

First, we assume that the reflectivity is split at the bandgap of the top cell ($\lambda_{step} = \lambda_{BG}$), and assume *r=1*. In that case for both spectral splitter configurations the absorption in the Si bottom cell is identical to the one in the thermodynamic limit as no light below the top cell bandgap energy is transmitted to the bottom cell. Figure 3b shows the tandem efficiencies for a 500 nm thick perovskite top cell, comparing the TTC limit with the planar and Lambertian spectral splitting limits. Figure 3c shows the corresponding top and bottom cell currents. For a 500 nm top cell the planar or Lambertian spectral splitters strongly improve the tandem efficiency. At 1.70 eV the Lambertian splitter increases the tandem efficiency by more than 5% to 42.9%. Interestingly, this is even beyond the TTC maximum of 42.0 % that occurs for a bandgap of 1.58 eV, reflecting the trend that the higher the absorption in the top cell, the closer the ultimate thermodynamic limit that assumes full absorption, which occurs at 1.73 eV, is achieved.

To further illustrate the benefit of the spectrum splitters, we show the TTC limit of a 1000 nm top cell in the same figure and compare the short-circuit current $J_{sc}$ of the top and bottom cells in in the lower panel. The top cell $J_{sc}$ for the 1000 nm thick cell and the 500 nm cell with planar spectral splitter is identical for all bandgaps as the top cell thickness is effectively doubled by the planar spectral splitter. The bottom cell current, however, is always lower for the 500 nm thick cell with spectrum splitter. This is because for the planar geometry above-bandgap light that is not absorbed in the top cell is transmitted into the Si bottom cell, while with the spectral splitter it is reflected and lost from the front side of the top cell. For the larger bandgaps, the efficiencies of the 1000 nm thick cell and the 500 nm cell with planar spectral splitter become equal as the identical top cell currents are the limiting factor in both tandems.

The benefit of the Lambertian splitter over the planar splitter is twofold: it further enhances absorption in the top cell due to the larger angular scattering range, and, as a result, less light is lost from the front of the cell. Consequently, the top cell current of the cell with Lambertian spectral splitter is consistently higher than that of the planar spectral splitter while the bottom cell current of both cases is equal.

**Splitting conditions**

Next, we investigate the influence on the efficiency of an offset of the step energy from the top cell bandgap energy as well as incomplete reflectance from the spectrum splitter. This is of interest because it has been found that for 4T tandem cells, especially for planar reflectors, the escape losses from the front side of the cell for light at energies just above the top cell bandgap can be detrimental for the overall tandem performance [19]. To avoid this, we shift the reflection spectrum to slightly higher energy, so that the non-absorbed spectral band just above the top-cell bandgap is transmitted to the Si bottom cell. Figures 4a and b show the efficiency gains/losses for a 500 nm top cell with a bandgap of 1.7 eV as a function of $\Delta E$ and *r*, with and without planar and Lambertian spectral splitter, respectively. In both cases the best result is found for the highest reflectivity of the spectral splitter, but the figure shows that also for non-ideal splitters with *r<1*, as they may be made experimentally, large efficiency gains are expected. For the planar splitter a maximum achievable efficiency enhancement above 4% is found; for the Lambertian one a gain of more than 5% is expected. In both cases, the optimum is found for a shift in the reflectance edge by about 10 meV above bandgap of the top cell. The maxima are marked with black stars in Figure 4a,b.

To further study the impact of the reflectance *r* we calculate the possible efficiency gain as a function of top cell bandgap for a Lambertian spectral splitter with 500 nm top cell thickness (Fig. 4c). In agreement with what is described above, efficiency gain is observed for the highest top cell bandgaps, while a loss is observed for the lowest gaps. However, we find that variation of *r* allows further optimization depending on the top-cell bandgap. An interesting subtlety occurs just at the cutoff energy 1.65 eV. Contrary to what was observed for the spectral splitter with *r=1* in Figure 3b, also for smaller bandgaps a spectral splitter geometry can be beneficial if the reflectivity is reduced. In Fig. 4c one can see that also here a Lambertian spectral splitter is beneficial and increases the efficiency by 2.7% absolute (marked with black star in Figure 4c), if the reflectance is set to be *r=0.63*. In detailed balance calculations, an equal increase can be reached by increasing the thickness of the top cell, however, in practice an increase in thickness leads to losses in the open-circuit voltage .

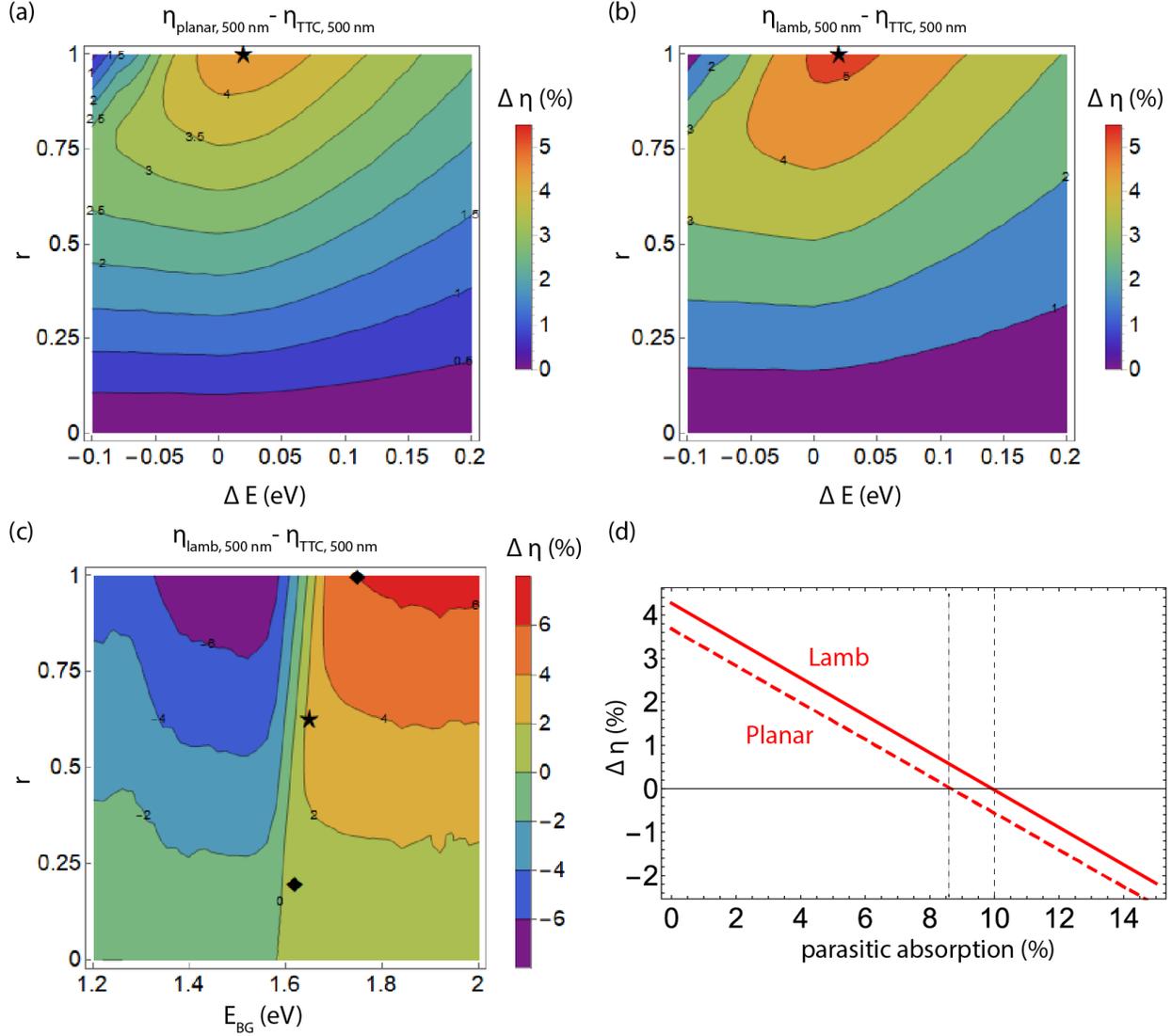

**Figure 4** Absolute efficiency gain of 2T tandem cells depending on reflectance of the spectral splitter, the step wavelength and parasitic absorption in the spectral splitter. Efficiency gain for tandem cells with planar (a) and Lambertian (b) spectral splitter and 500 nm thick perovskite top cell with $E_{BG}$=1.7 eV compared to TTC limit of the same cell, as function of *ΔE* and *r*. (c) Efficiency gain for tandem cell with Lambertian spectral splitter and 500 nm thick top cell as function of top cell bandgap and reflectance *r* with *ΔE*=0.01 eV. The star indicates the ideal spectral splitting condition for maximal efficiency gain for a tandem cell with a 500 nm thick perovskite top cell with $E_{BG}$=1.65 eV. The diamonds mark the position of maximum gain for two examples of realistic perovskite-silicon cells with $E_{BG}$=1.62 eV and $E_{BG}$=1.77 eV. (d) Efficiency gain for a tandem cell with Lambertian (solid red line) and planar (dashed red line) spectral splitter and 500 nm thick top cell as function of parasitic absorption in the spectral splitter.

We note that all calculations so far were based on an idealized set of assumptions. To get an idea of what efficiency enhancement can be expected in the Lambertian case for realistic materials, we applied detailed-balance calculations using data from selected record cells [33]. As a first estimate, we took a state-of-the-art perovskite/Si tandem cell with a top-cell bandgap of 1.62 eV. The corresponding EQE-spectra and thickness for the top cell layer were used to approximate the absorption coefficient around the bandgap, by assuming unity internal quantum efficiency. The resulting absorption coefficient was used to investigate the potential performance of a Lambertian spectral splitter for a 500 nm thick perovskite layer on top of Si. Figure 4c contains markers (diamonds) representing this comparison. In the case of the 1.62 eV perovskite, it turns out that an efficiency enhancement of 0.30% absolute could be achieved (in the detailed-balance limit) for an ideal reflectance of $r$=0.2.

So far, this analysis assumed detailed-balance-derived open-circuit voltage ($V_{oc}$) and fill-factor (FF) values. To get a more accurate estimate for realistic conditions, we approximate the tandem cell as a non-ideal diode under illumination described by $J(V) = J_{SC} - J_0 * (\exp\left(\frac{q*V}{n*k*T}\right) - 1)$, where the overall current density $J(V)$ depends on the short-circuit current density $J_{SC}$ and the voltage dependent diode current density term with the voltage $V$, the reverse saturation current density $J_0$, an ideality factor $n$, Boltzmann's constant $k$ and the temperature $T$=300K. Using this equation, we can fit the effective recombination current $J_{0,tandem}$ and ideality factor $n_{tandem}$ of the full tandem device from the cell's IV-curve. The $V_{OC}$ was calculated by using the $J_{SC}$ for the Lambertian case as derived above in the detailed-balance calculations, plugging it into the non-ideal diode equation and solving the equation for $J(V)$=0. Furthermore, we account for the local FF minimum around the current matching condition [34] by applying the same relative FF-loss that is observed in the detailed-balance analysis above on the listed FF value of the experimental cell. This comparison yields 0.26% absolute efficiency enhancement ($r$=0.2), just slightly lower compared to the initial estimate that was based on only adjusting the absorption coefficient to the experimental value.

Considering the two estimated values are rather close to each other, we feel confident in doing estimations for cases for which the nature of the available data only allows an estimate according to the first method. We take the top cell from the current perovskite/perovskite record tandem cell (1.77 eV) [33] and determine the absorption coefficient. For the imagined case of such a top cell on top of a Si bottom cell, this analysis yields an efficiency enhancement of 2.8% absolute at a reflectance of $r$=1, highlighting that the significance of a Lambertian spectral splitter increases with the top cell bandgap. We note that the efficiency gains according to this estimate are less than what Figure 4c suggests (upper diamond), and this is attributed to the sharper onset of the EQE-derived absorption coefficients compared to the reference absorption coefficient that was used for the modelling of Figure 4.

Finally, an important parameter in the experimental realization of a spectral splitter are the losses that such a layer could introduce into the system. Figure 4d shows the efficiency gain/loss as a function of absorption, assuming spectrally flat parasitic absorption for a 500 nm thick cell. We find that for absorption up to 9/11% the planar/Lambertian spectral splitter is still beneficial.

**Conclusion**

In this work, we derive the detailed-balance efficiency limit of two-terminal tandem solar cells with a perovskite top cell and a Si bottom cell, considering the realistic incomplete absorption conditions for the perovskite top cell. We calculate the theoretically possible efficiency gain due to the introduction of a spectral splitter in between the top and bottom cell. For top cells with bandgaps above 1.7 eV, a spectral splitter strongly enhances light absorption in the top cell, leading up to 5-6% absolute efficiency gain in the thermodynamic limit for a 500-nm thick top cell. Using experimental parameters of realistic cells, we predict an efficiency gain for a practical perovskite-tandem cell with Lambertian spectral splitter of 2.8% for a top cell with a bandgap energy of 1.77 eV and unity spectrum splitter reflectivity. In optimizing the reflectivity of the spectral splitter we find small subtleties in the 1.6-1.7 eV bandgap range, due to unwanted outscattering of light from the front side of the cell. The effect of parasitic absorption that will occur in experimental spectral splitters is also derived. Overall, our work shows there is a bright perspective for the integration of spectral splitters in perovskite-Si tandem solar cells, even if unity reflectivity cannot be achieved experimentally.

**Acknowledgement**